\documentclass[aps,prl,twocolumn,showpacs]{revtex4-1}
\usepackage{graphicx}
\usepackage{bm}
\usepackage{wasysym}
\usepackage{setspace}

\begin{document}

\title{Deconfined Fractionally Charged Excitation in Any Dimensions}
\author{Chyh-Hong Chern}
\email{chchern@ntu.edu.tw}
\author{Po-Hao Huang}
\affiliation{Department of Physics and Center for Theoretical Sciences, National Taiwan University, Taipei 10617, Taiwan}
\date{\today}
\begin{abstract}
An exact incompressible quantum liquid is constructed at the filling factor $1/m^2$ in square lattice.  It supports deconfined fractionally charged excitation.  At the filling factor $1/m^2$, the excitation has fractional charge $\pm e/m^2$, where $e$ is the electric charge.   This model can be easily generalized to the integer lattice in any $D$ dimensions, where the charge of excitations becomes $\pm e/m^D$. 
 \end{abstract}

\maketitle


Quantum number fractionalization is one of the most intriguing phenomena in physics.  The most astonishing discovery of the fractional excitation is arguably the Laughlin quasiparticle in the fractional quantum Hall effect (FQHE).  Laughlin quasiparticle has charge $e/q$, where $e$ is the electric charge and $q$ is taken to be any odd integer.  Different from quarks, it is indisputably a fractional excitation because 1) it is in the deconfined phase and 2) it is a new degree of freedom in addition to the ground state which is an electronic quantum liquid.

In one dimension, quantum number fractionalization is a robust effect.  There are many models in one dimension exhibiting the fractionalization of charge or spin in the excitation spectrum.  On the contrary, it is difficult to be found in theoretical models and in experiments in the dimensions greater than one, although there is no theorem to restrict its existence in higher dimensions.  Whether or not deconfined fractionalization can \emph{in principle} occur in any dimensions is still an open question.  The understanding to this problem is still primitive, neither is there a general theoretical construction.  In this Letter, we answer the question positively by providing, to our best knowledge, the first microscopic model to host its existence in the integer lattice in any dimensions.  Specifically, the two-dimensional case in the square lattice will be worked out in detail.  The higher dimensional generalization can be systematically constructed.  It is the first theoretical model to host the exact ground state of incompressible quantum liquid in any dimensional integer lattice.

The model we consider here was originally proposed by one of us (Chern) and Lee \cite{chern2007prl}, where a lattice pair-hopping model of incompressible quantum liquid is constructed in the $(n-1)$-dimensional triangular lattice.  It is basically the generalization of the fractional quantum Hall effect in CP$^{n-1}$, which is the $2(n-1)$-dimensional complex projective space.   In ref.\cite{chern2007prl}, we constructed the generalized Laughlin wavefunction in CP$^{n-1}$ and found the Hamiltonian to host it as the \emph{unique} ground state which is well separated from the excitations by a finite energy gap.  The "lattice" is the weight space of the coherent states of SU($n$), the isometry group of CP$^{n-1}$, in the $(p,0,...,0)$ representation, where $p$ is an integer.  Without losing generality, in the following we start with the simplest generalized case, namely CP$^2$.  After briefly reviewing the FQHE in CP$^2$ and taking the flat space limit from CP$^2$ to R$^4$, we shall illustrate that the incompressible quantum liquid can be constructed in the square lattice with \emph{periodic} boundary.  Furthermore, we shall show that the fractionally charged excitation, appearing as a dislocation, is in the deconfined phase.  Most importantly, our theory can be generalized to higher dimensions easily.

The Laughlin wavefunction in CP$^2$ can be written by
\begin{eqnarray}
\Psi^m = \left| \begin{array}{ccccc} u_1^p & u_1^{p-1}v_1 & . & .
& w_1^p \\ u_2^p & u_2^{p-1}v_2 & . & . & w_2^p \\ . & . & . & . &
.  \\ . & . & . & . & . \\ u_N^p & u_N^{p-1}v_N & . & . & w_N^p
\end{array} \right|^m, \label{Eq:su3_FQHE}
\end{eqnarray}
where $u,v,w$ form the SU(3) fundamental triplet, $N$ is the number of particles, $m$ is any odd integer, and together with $p$ they label the SU(3) $(mp,0)$ representation of which the weight space spans a two-dimensional triangular lattice with \emph{open} boundary.  The relation between $N$ and $p$ is given by $N=d(p,0)$, where $d(p,0)=(p+1)(p+2)/2$ the dimension of the SU(3) $(p,0)$ representation.  For general $m$, the filling factor $\nu = d(p,0)/d(mp,0)$, which equals to $1/m^2$ in the thermodynamic limit.

For convenience, we consider $m=3$ in the following.  It will be soon clear that it is not hard to generalize to any odd $m$.  In the flat space limit, $(u,v,w)$ becomes $(1, z^1, z^2)$, where $z^1=x^1+ix^2$ and $z^2=x^3+ix^4$ parametrizing the R$^4$ space.  In this limit, Eq.(\ref{Eq:su3_FQHE}) can be written by
\begin{eqnarray}
\Psi_{R^4}^3 = \left| \begin{array}{ccccc} 1 & z^1_1 & . & .
& (z^2_1)^p \\ 1 & z^1_2 & . & . & (z^2_2)^p \\ . & . & . & . &
.  \\ . & . & . & . & . \\ 1 & z^1_N & . & . & (z^2_N)^p
\end{array} \right|^3, \label{Eq:R4_FQHE}
\end{eqnarray}
The R$^2$ version of Eq.~(\ref{Eq:R4_FQHE}), the Laughlin wavefunction, is well known to be the unique ground state of the Trugman-Kivelson Hamiltonian~\cite{trugman1985PRB}.  To generalize, we shall show that Eq.~(\ref{Eq:R4_FQHE}) is the unique ground state of the following Hamiltonian,
\begin{eqnarray}
H=\sum_{i<j}\nabla^2_{ij}\delta(\vec{r}_{ij}), \label{Eq:R4_Hami}
\end{eqnarray}
where $\vec{r}_{ij}=(x^1_i-x^1_j,x^2_i-x^2_j,x^3_i-x^3_j,x^4_i-x^4_j)$, $\delta(\vec{r}_{ij})$ is the $\delta$-function, and $\nabla_{ij}$ is the four-dimensional differential operator.  To show Eq.~(\ref{Eq:R4_FQHE}) to be the ground state of Eq.~(\ref{Eq:R4_Hami}), we first observe that for any pair $i$ and $j$, Eq.~(\ref{Eq:R4_FQHE}) can be expressed by
\begin{eqnarray}
\Psi_{R^4}^3 = [(z^1_i-z^1_j)F_1+(z^2_i-z^2_j)F_2]^3, \label{Eq:R4_FQHE_m3}
\end{eqnarray}
where $F_{1,2}$ contain the symmetric part, including the even power of $(z^{1,2}_i-z^{1,2}_j)$.  It is the property of the antisymmetry of the many-body wavefunction.  Eq.~(\ref{Eq:R4_Hami}) implies that the zero energy state should have $(z^{1}_i-z^{1}_j)^{\alpha}(z^{2}_i-z^{2}_j)^{\beta}$ where $\alpha+\beta=$ odd integers $\ge 3$.  Eq.~(\ref{Eq:R4_FQHE_m3}) can be easily checked to possess this property.  Since Eq.~(\ref{Eq:R4_Hami}), describing the short-ranged repulsive interaction, is positive definite, Eq.~(\ref{Eq:R4_FQHE}), the Laughlin wavefunction in R$^4$, is the ground state of Eq.~(\ref{Eq:R4_Hami}).  Moreover, using the similar argument in ref.~\cite{chern2007prl, chern2010prb}, the uniqueness of the ground state can be easily proved.  We further note that the finiteness of the excitation gap of Eq.~(\ref{Eq:R4_Hami}) is already shown in ref.~\cite{chern2007prl, chern2010prb}, since the calculation was performed in the R$^4$ limit.  Without elaborating too much technical detail, we remark that Eq.~(\ref{Eq:R4_Hami}) is actually the flat space limit of the pseudo-potential Hamiltonian in CP$^2$ given in ref.~\cite{chern2007prl}.

Next, let us proceed to transform the theory to a lattice model.  The dynamics of QHE in R$^4$ in the lowest Landau level can be viewed as if there are two-independent non-commutative planes ~\cite{bernevig2002ANN, chern2010prb}.  One can choose them to be the $x^1$-$x^2$ plane and the $x^3$-$x^4$ plane.  By applying the periodic boundary condition to the $x^2$ and $x^4$ directions, the momentum in these two directions is quantized.  Then, the single-particle wavefunction can be written by ~\cite{Lee2004PRL, Seidel2005}
\begin{eqnarray}
\phi_{mn} =\frac{1}{l_B\sqrt{\pi L_1L_2}}e^{i\frac{2m\pi x^2}{L_1}}\exp(-\frac{(x^1-a_1m)^2}{2l^2_B})\nonumber\\ \times e^{i\frac{2n\pi x^4}{L_2}}\exp(-\frac{(x^3-a_2n)^2}{2l^2_B}), \label{Eq:2T2_orbital}
\end{eqnarray}
where $L_{1,2}$ are the linear dimension in the $x^{1,3}$ directions, $m$ and $n$ are the quantum numbers of the momentum in the $x_2$ and $x_4$ directions respectively, $l_B=\sqrt{\hbar c/eB}$ is the magnetic length, and $a_{1,2}=\frac{2\pi l^2_B}{L_{1,2}}$.  These single-particle orbitals are Gaussian wave packages which localize in the $x_1$ and $x_3$ directions denoting by $m$ and $n$.  It defines a two-dimensional lattice with lattice constants $a_{1,2}$, and $m$ and $n$ are the label of sites.  Since $L_{1,2}$ are finite, it automatically sets the upper bounds of $m$ and $n$, that are $L_{1}/a_1$ and $L_2/a_2$ respectively.  Furthermore, $\phi_{mn}$ for different $m$ and $n$ are orthogonal.  Therefore, $\phi_{mn}$ can be viewed as Wannier basis in this system.

In order to restore the translational symmetry, periodic boundary condition has to be applied in the $x^1$ and $x^3$ directions.  In this case, $\phi_{mn}$ is modified to be
\begin{eqnarray}
&&\phi_{mn} = \nonumber\\&&\frac{1}{l_B\sqrt{\pi L_1L_2}}\sum_{s,t}^{\text{int.}}e^{i(\frac{2m\pi}{L_1}+\frac{2s\pi}{a_1})x^2}\!\exp(-\frac{(x^1\!-\!a_1m\!-\!sL_1)^2}{2l^2_B})\nonumber\\&& \times e^{i(\frac{2n\pi}{L_2}+\frac{2t\pi}{a_2})x^4}\exp(-\frac{(x^3-a_2n-tL_2)^2}{2l^2_B}) \label{Eq:periodicorbital}
\end{eqnarray}
Now, defining the field operator on this Wannier basis $\Psi(\vec{x})=\sum_{mn} c_{mn}\phi_{mn}(\vec{x})$, where $c_{mn}$ is the annihilation operator of electron at the site $(m,n)$, the second quantization of Eq.~(\ref{Eq:R4_Hami}) can be obtained easily as follows
\begin{eqnarray}
&\!\!\!\!\!&H\!\!=\!\!\frac{2\kappa^4}{\pi^3 l_B^6}\!\!\!\sum_{1\leq p\leq\frac{L_1}{a_1}\!+\!\frac{1}{2}}\sum_{1\leq q\leq\frac{L_2}{a_2}\!+\!\frac{1}{2}}\sum_{0 \leq |l|,|l'|\leq \frac{L_1}{2a_1}}\sum_{0\leq |d|,|d'|\leq \frac{L_2}{2a_2}}\nonumber\\&\!\!&f(l,l';d,d')c^\dag_{p+l',q+d'}c^\dag_{p-l',q-d'}c_{p+l,q+d}c_{p-l,q-d}  \label{Eq:2ndquantization}
\end{eqnarray}
where
\begin{eqnarray}
&&f(l,l';d,d')=\nonumber\\&&\sum_{n,n'}^{int.}\sum_{j,j'}^{int.}[(l+n\frac{L_1}{a_1})(l'+n'\frac{L_1}{a_1})+(d+j\frac{L_2}{a_2})(d'+j'\frac{L_2}{a_2})]\nonumber\\&&e^{-\kappa_1^2[(l+n\frac{L_1}{a_1})^2+(l'+n'\frac{L_1}{a_1})^2]}e^{-\kappa_2^2[(d+j\frac{L_2}{a_2})^2+(d'+j'\frac{L_2}{a_2})^2]} \nonumber
\end{eqnarray}
and $\kappa_i = a_i/l_B$.  If the hopping ranges $1/\kappa_i$ are measured in the unit of lattice constants, we found the hopping range $a_i/\kappa_i=l_B$, which are the same in both directions.  Furthermore,  one can easily check that the hopping integral $f(l,l';d,d')$ in Eq.~(\ref{Eq:2ndquantization}) has a p4m group symmetry, known as the symmetry of the square lattice.  It is certain that $a_1=a_2=a$ manifests the square lattice.  Therefore, Eq.~(\ref{Eq:2ndquantization}) describes a lattice model in the two-dimensional square lattice with periodic boundary.  

Eq.~(\ref{Eq:2ndquantization}) is the two-dimensional generalization of the model proposed by Seidel et al.~\cite{Seidel2005}.  It is a pair hopping model with center-of-mass position conservation.  In other words, it commutes with the following operators simultaneously:
\begin{eqnarray}
U_x=e^{ i\frac{2\pi a}{L_1}\sum_{i}xc^\dag_ic_i}, \ \ U_y=e^{ i\frac{2\pi a}{L_2}\sum_{i}yc^\dag_ic_i},
\end{eqnarray}
where $i$ runs over all lattice sites which are labelled by two integers $(x,y)$.  Since the lattice has a periodic boundary, the system is translationally invariant.  Namely, Eq.~(\ref{Eq:2ndquantization}) also commutes with the translational operator $T_x$ and $T_y$ which translate the system by one lattice constant $a$ in $x$ and $y$ directions respectively.  However, $U_i$ and $T_i$ do not mutually commute.  At filling $1/m^2$, they satisfy the following algebra
\begin{eqnarray}
&&T_xU_xT^\dag_x=U_xe^{i\frac{2\pi l_2}{m}}, \ \ T_yU_yT^\dag_y=U_ye^{i\frac{2\pi l_1}{m}} \nonumber \\&&
[T_i,U_j]=0 \ \ \text{for} \ i\ne j \label{Eq:UT}
\end{eqnarray}
where $L_i=ml_ia$ and $l_i$ are integers chosen to be prime to $m$.  Eq.~(\ref{Eq:UT}) implies that the ground states cannot be simultaneously the eigenstates of $T$ and $U$.  One can choose the ground state to be the eigenstate of $U$ satisfying $U_x|e^{i\phi_x},e^{i\phi_y},E>=e^{i\phi_x}|e^{i\phi_x},e^{i\phi_y},E>$ and $U_y|e^{i\phi_x},e^{i\phi_y},E>=e^{i\phi_y}|e^{i\phi_x},e^{i\phi_y},E>$.  Then, $T_x^pT_y^q|e^{i\phi_x},e^{i\phi_y},E>$ are the orthogonal ground states for $p$ and $q$ to be the integer from 0 to $m-1$.  In fact, the algebra in Eq.~(\ref{Eq:UT}) imply that the whole spectrum has at least $m^2$-fold degeneracy~\cite{oshikawa2000PRL}.

\begin{figure}[htb]
\includegraphics[width=0.5\textwidth]{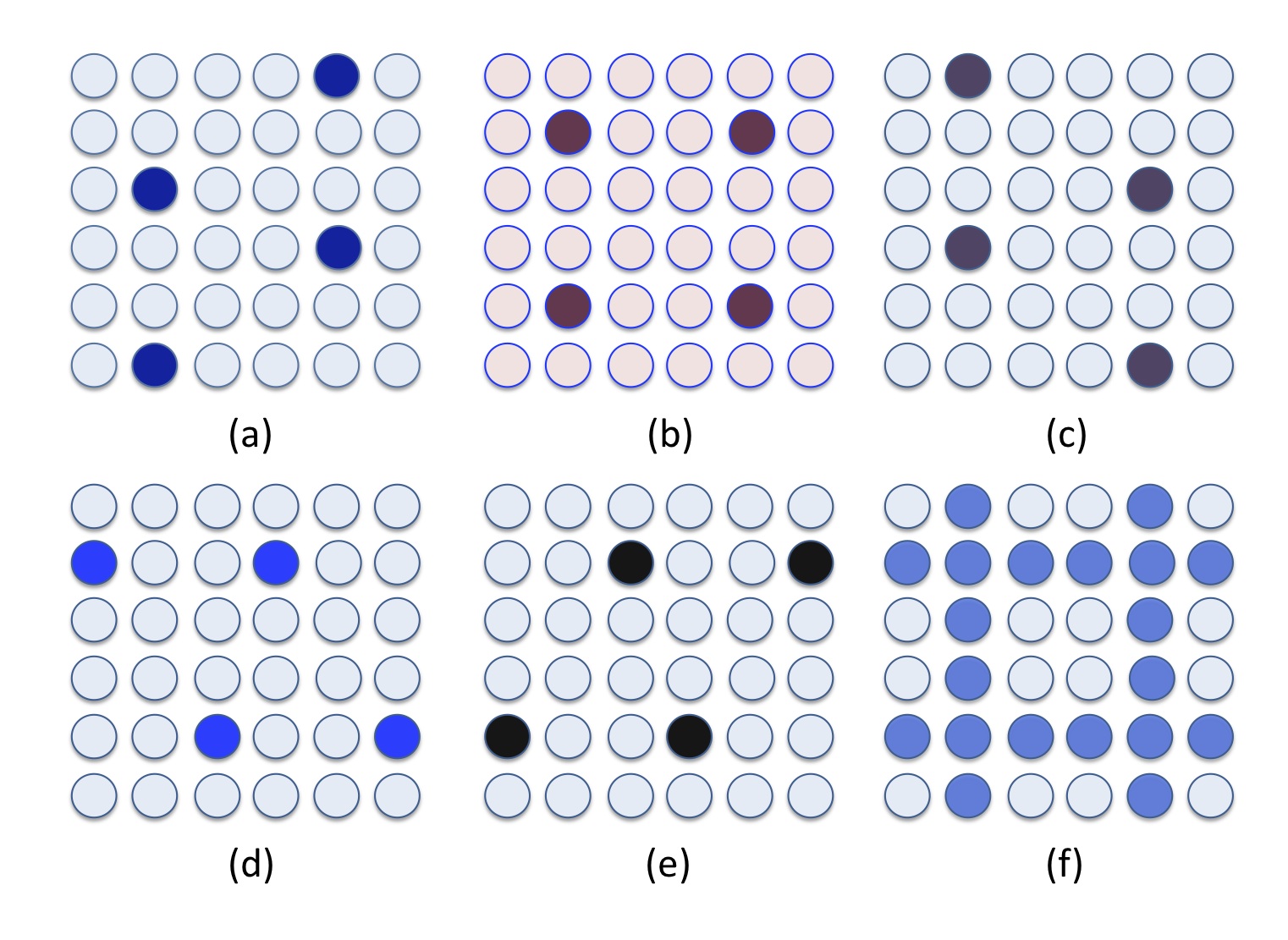}
\caption{(Color online) (a) to (e) the ground states in the $\kappa >> 1$ limit in the $N=4$ case.  These five states have the same center-of-mass position.  (f) the state of the equally-weighted linear superposition of the states from (a) to (e).}\label{Fig:groundstate}
\end{figure}

Now, let us analyze the ground state of Eq.~(\ref{Eq:2ndquantization}) of $m=3$.  For $\kappa << 1$,  which can be achieved by taking $l_B << L$, the Wannier basis $\phi_{mn}$ highly overlap in space, since the lattice constant $a$ decreases faster than $l_B$.    The ground state corresponds to the FQH states of Eq.~(\ref{Eq:R4_FQHE}).  Different from Eq.~(\ref{Eq:R4_FQHE}), the ground state of Eq.~(\ref{Eq:2ndquantization}) is at least 9-fold degeneracy.  The generate ground states are actually distinguished by the center-of-mass positions.   As we shall show later, the ground state for small $\kappa$ indeed exhibits vanishing local order parameters.  For $\kappa > > 1$, one can expand the Hamiltonian in terms of $e^{-\kappa^2/2}$.  We found Eq.~(\ref{Eq:2ndquantization}) can be well approximated by
\begin{eqnarray}
&H&=\!\sum_{i,j}[f(\frac{1}{2},\frac{1}{2};0,0)n_{i+1,j}n_{i,j}\!\!+\!\!f(0,0;\frac{1}{2},\frac{1}{2})n_{i,j+1}n_{i,j}\!+\nonumber\\ &\!&\!f(\frac{1}{2},\frac{1}{2};\frac{1}{2},\frac{1}{2})n_{i+1,j+1}n_{i,j}+f(1,1;0,0)n_{i+2,j}n_{i,j\!}+\nonumber\\&\!&\!f(0,0;1,1)n_{i,j+2}n_{i,j}+f(1,1;\frac{1}{2},\frac{1}{2})n_{i+2,j+1}n_{i,j}\!+\nonumber\\&\!&\!f(\frac{1}{2},\frac{1}{2};1,1)n_{i+1,j+2}n_{i,j}\!+\!f(1,1;1,1)n_{i+2,j+2}n_{i,j}]\!+\nonumber\\&\!&\!O(e^{-\frac{3}{2}\kappa^2},e^{-\frac{5}{2}\kappa^2},e^{-3\kappa^2},e^{-\frac{7}{2}\kappa^2}), \label{Eq:largekappa}
\end{eqnarray}
where $n_i=c_i^\dag c_i$ and $O(e^{-\frac{3}{2}\kappa^2},e^{-\frac{5}{2}\kappa^2},e^{-3\kappa^2},e^{-\frac{7}{2}\kappa^2})$ denotes the positive energy for a pair gained by making the separation distance less than or equal to $\sqrt{8}a$ before or after hopping.  Because $f>0$, Eq.~(\ref{Eq:largekappa}) stabilizes a ground state of charge density wave (CDW) such that \emph{no pair is separated less than or equal to $\sqrt{8}a$}.  We plot the ground states for $N=4$ in Fig.~(\ref{Fig:groundstate}).  Given the center-of-mass position fixed, there are more ground state degeneracy in addition to the 9-fold degeneracy.  For $N=2$, the degeneracy is 3, for $N=4$ there are 5 shown in Fig.~(\ref{Fig:groundstate}a) to Fig.~(\ref{Fig:groundstate}e), and for $N=6$ there are 15, and therefore the total ground state degeneracy is 27, 45, and 45 respectively.  In general, for $N>>1$, the degeneracy for fixed center-of-mass position is ${\bf M}=3^{l_1-1}+3^{l_2-1}-1$ with $l_1$ and $l_2$ to be indivisible by 3, and the total ground state degeneracy is 9{\bf M}.  We stress that the result of the additional degeneracy is not an artifact of the truncated Hamiltonian.  We performed the exact diagonalization calculation in the $N=2,4$, and 6 cases.  We not only reproduce the analytical results given above but also obtain that the ground state degeneracy is \emph{independent} of $\kappa$.  In the $\kappa >> 1$ limit, the uniform ground state can be obtained by the equally-weighted linear superposition of the degenerate ground states for a fixed center-of-mass position.  The one in the $N=4$ case is plotted in Fig.~(\ref{Fig:groundstate}f).

\begin{figure}[htb]
\includegraphics[width=0.5\textwidth]{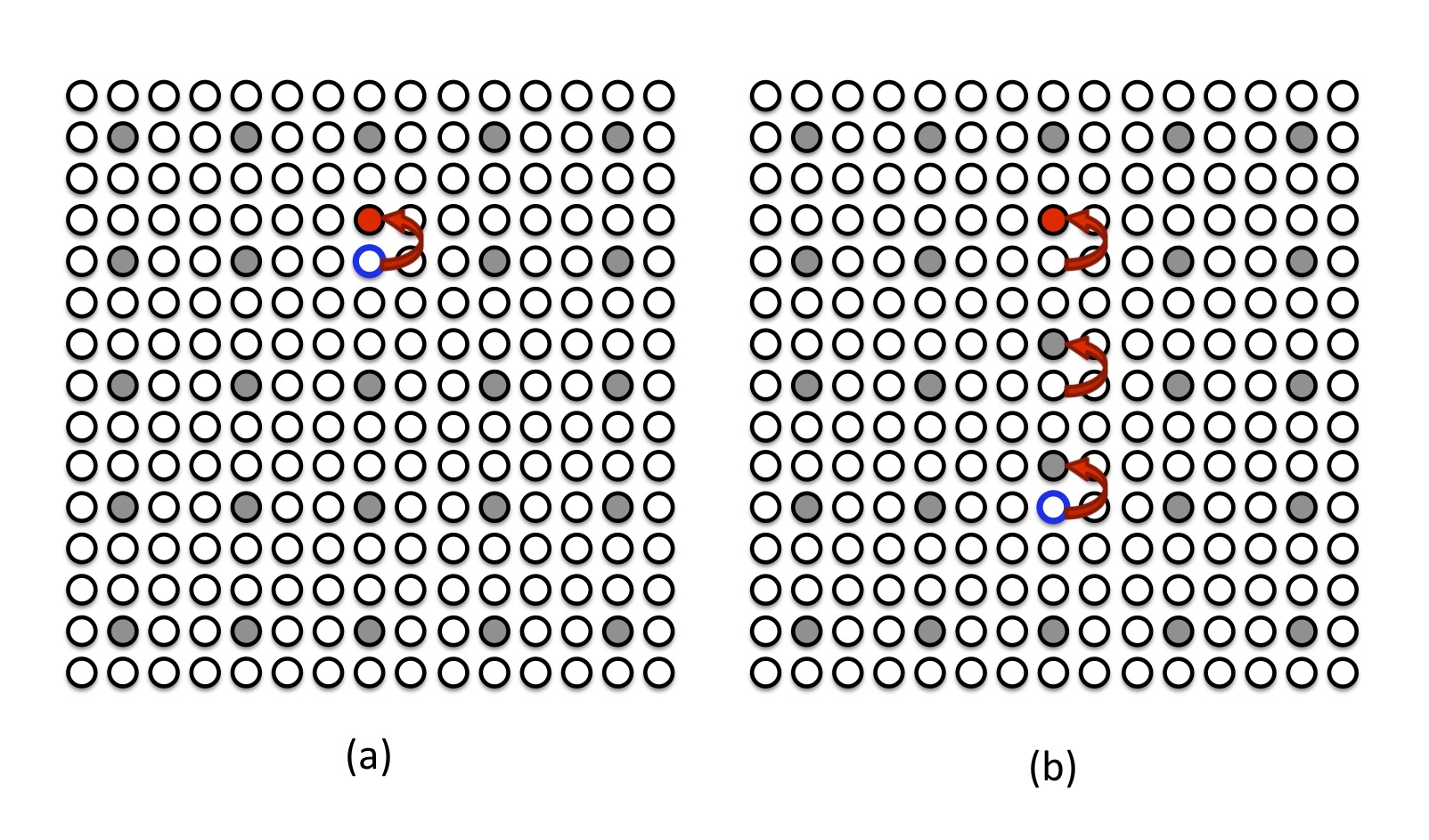}
\caption{(Color online) (a) the quasiparticle-quasihole pair can be created by shifting an electron up.  The red curve arrow denotes the hopping of electron. (b) an example of the string excitation.  The energy of the string excitation is independent of the length of the string, indicating deconfinement.  The red ball denotes the location of the quasipaticle, and the blue circle is the one of the quasihole.}\label{Fig:excitation}
\end{figure}
From Eq.(\ref{Eq:largekappa}), it can be clearly seen that a excitation can be created if the distance between any two particles is equal to or smaller than $\sqrt{8}a$.   We can simply move an electron up by one site as shown in Fig.~(\ref{Fig:excitation}a).  Then, one can make a string excitation by moving more electrons up as shown in Fig.~(\ref{Fig:excitation}b).  It turns that the energy of the excitations in Fig.~(\ref{Fig:excitation}a) and Fig.~(\ref{Fig:excitation}b) is actually the same, because creating a string excitation does not make more pairs to be separated less than or equal to  $\sqrt{8}a$.  Therefore, the fractional excitations occur at both ends of the string.  One end corresponds to the quasipaticle and the other end is the quasihole.  The creation of quasiparticle and quasihole pair costs finite energy $\sim e^{-2\kappa^2}$.  However, it does not cost energy to separate them.  In other words, the quasiparticles and quasiholes are in the \emph{deconfined} phase.

\begin{figure}[htb]
\includegraphics[width=0.5\textwidth]{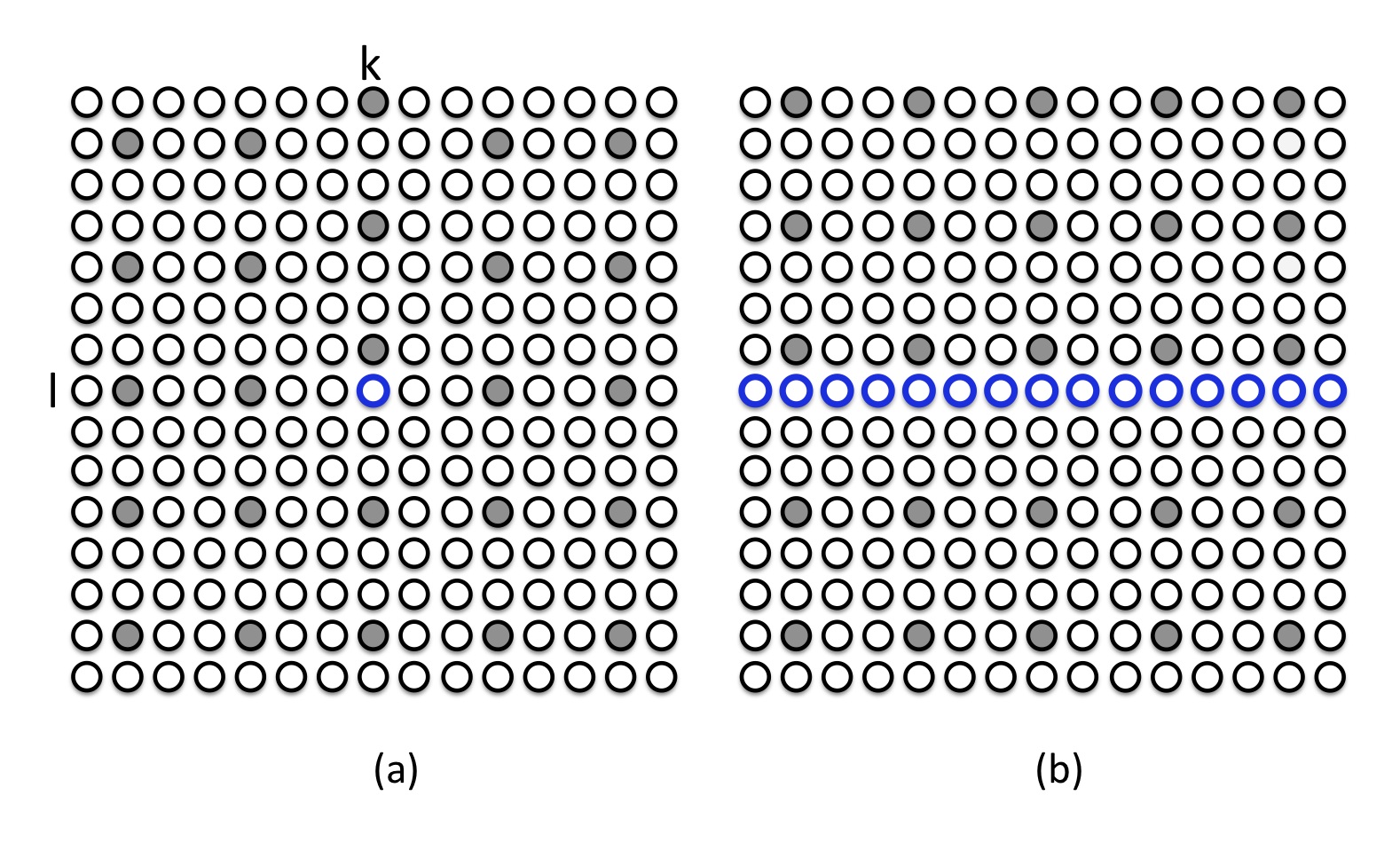}
\caption{(Color online) (a) a quaishole is a dislocation locating at $(k,l)$. (b) a domain wall formed by the quasiholes}\label{Fig:charge}
\end{figure}
The charge of the quasihole can be computed as the following. In a $L_x\times L_y$ lattice with open boundary condition, let us create a quasihole at $(k,l)$ shown in Fig.~(\ref{Fig:charge}a).  It is nothing but inserting an empty site and create a dislocation in the system of CDW.  Let us assume that its charge is $e^*$.  One can add more empty sites at row $l$ so that a domain wall is created shown in Fig.~(\ref{Fig:charge}b).  The charge of the domain wall is thus $e^*L_x$.  Now, one can create one more domain wall by adding $L_x$ sites.  If we add three domain walls, it is effectively equivalent to remove $L_x/3$ electrons.  Since three domain walls has charge $3e^*L_x$, we obtain $e^*= e/9$, where $e$ is the electric charge.  For general $m$, the fractional excitation has $e^*=e/m^2$.  It is easy to generalize to $D$-dimensional integer lattice.  Quasiparticle/quasihole is a dislocation behaving like a point-like particle in the D-dimensional lattice, where the fractional charge is given by $e/m^{D}$.  It is not surprising that the charge depends on the dimensionality, because the fractional excitation is a point-like particle.

\begin{figure}[htb]
\includegraphics[width=0.5\textwidth]{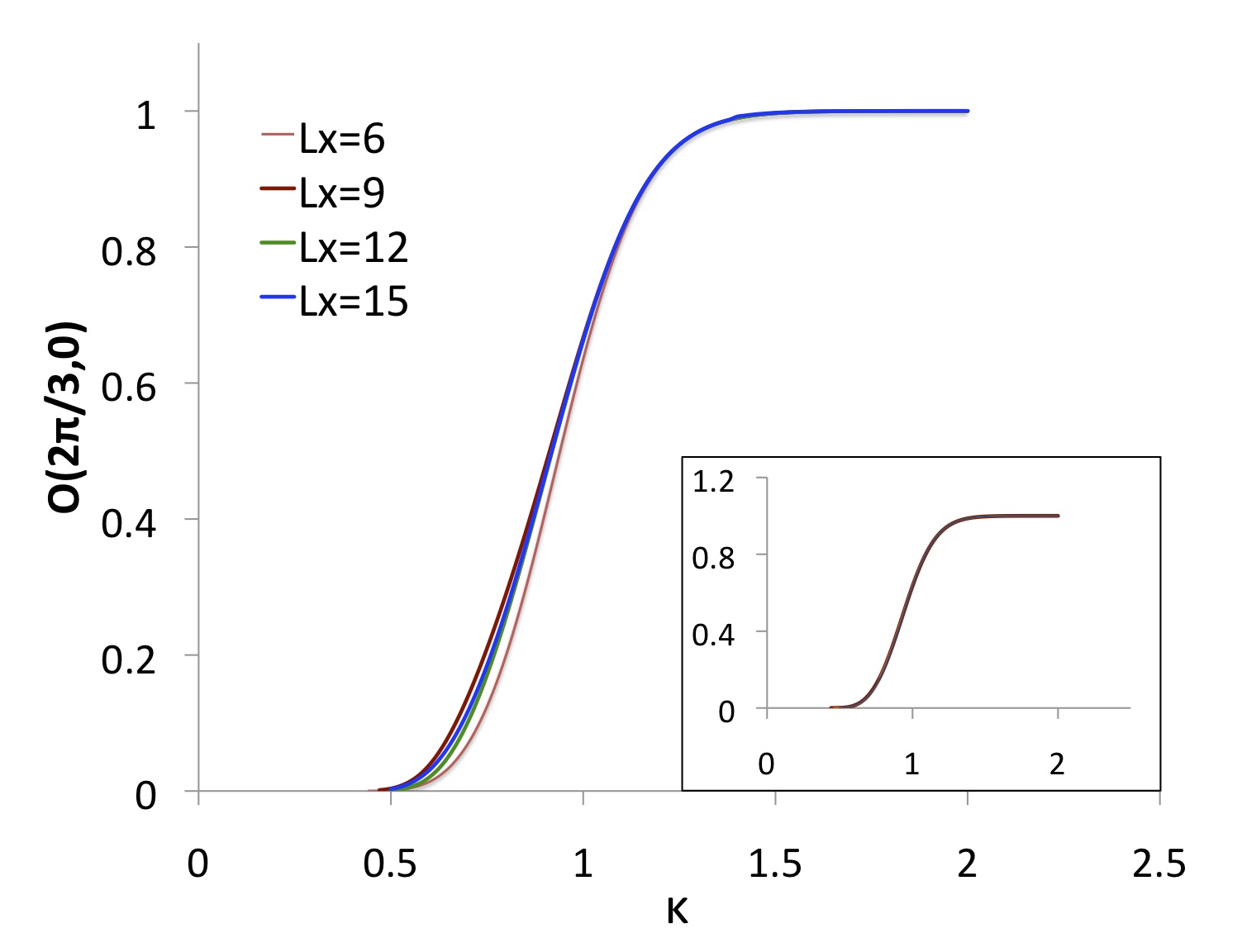}
\caption{(Color online) In the inset, the scaling of $\mathcal{O}_{\frac{2\pi}{3},0}$ with $L_y$ is calculated in the following cases $L_x\times L_y=6\times 3, 6\times6$, and $6\times 9$.  The convergence is so fast that they are hardly distinguishable.  Taking the value of $\mathcal{O}_{\frac{2\pi}{3},0}$ in the infinite $L_y$ limit, the scaling of $\mathcal{O}_{\frac{2\pi}{3},0}$ with $L_x$ is computed for $L_x=6,9,12$, and 15.}\label{Fig:order}
\end{figure}

We believe that the FQH state at $\kappa << 1$ can be adiabatically connected to the charge-density-wave state at $\kappa >>1$ without phase transition.  In one dimension, Seidel et al. has proven it true~\cite{Seidel2005}.  Here, we perform three calculations to support this conjecture. First, we checked that the ground state degeneracy is independent of $\kappa$.  We have performed the exact diagonalization to prove it true in the systems of 18 and 36 lattice sites.  Second, we compute the following order parameter as function of $\kappa$
\begin{eqnarray}
&&\mathcal{O}_{(\frac{2\pi}{3},0)} = \frac{1}{N}\sum_{m,n}e^{i\frac{2\pi m}{3}}c^\dag_{mn}c_{mn}. 
\end{eqnarray} 
In the finite size system, $\mathcal{O}_{(\frac{2\pi}{3},0)}$ is a function of $L_x$ and $L_y$.  In the inset of Fig.~(\ref{Fig:order}), we fix $L_x$ and show the scaling of $\mathcal{O}_{(\frac{2\pi}{3},0)}$ with $L_y$.  Then, taking the infinite $L_y$ limit, we plot the scaling of $\mathcal{O}_{(\frac{2\pi}{3},0)}$ with $L_x$.  We found that the scaling of $\mathcal{O}_{(\frac{2\pi}{3},0)}$ with $L_x$ is almost identical to the 1D case given by Seidel et al.~\cite{Seidel2005}.  The scaling shows very nice convergence.  The smooth change of the order parameter indicates that there is no phase transition between the $\kappa << 1$ and $\kappa >> 1$ phases.  Third, although it is not shown here, we also compute the energy gap for the system of 36 lattice sites.  We found that the energy gap actually increase at around $\kappa = 0.5$ where the order parameter is about to deviate from zero observed by naked eyes and remains gapped in the whole region of crossover.  We note that similar to the 1D case the order parameter actually decreases exponentially below $\kappa < 0.5$.  With those three calculations mentioned above, we believe that the incompressible quantum liquid at $\kappa << 1$ hosts the excitation of the charge $e^*=e/m^2$.

In summary, we show the existence of the fractionally charged excitation in a lattice hopping model in the square lattice.  Our construction can be systematically generalized to higher dimensional square lattice, namely the integer lattice.  The current theory answers positively that the fractionally charged excitation can \emph{in principle} exist in any dimension.  Our model is the first one to host the ground state of incompressible quantum liquid in the integer lattice.


CHC deeply appreciates the stimulating discussion with Dung-Hai Lee.  We express sincere gratitude to Hong-Shi Lee to scrutinize our calculation details.  This work is supported by NSC 97-2112-M-002-027-MY3 of Taiwan.



\begin{thebibliography}{1}%
\makeatletter
\providecommand \@ifxundefined [1]{%
 \ifx #1\undefined \expandafter \@firstoftwo
 \else \expandafter \@secondoftwo
\fi
}%
\providecommand \@ifnum [1]{%
 \ifnum #1\expandafter \@firstoftwo
 \else \expandafter \@secondoftwo
\fi
}%
\providecommand \enquote [1]{``#1''}%
\providecommand \bibnamefont  [1]{#1}%
\providecommand \bibfnamefont [1]{#1}%
\providecommand \citenamefont [1]{#1}%
\providecommand\href[0]{\@sanitize\@href}%
\providecommand\@href[1]{\endgroup\@@startlink{#1}\endgroup\@@href}%
\providecommand\@@href[1]{#1\@@endlink}%
\providecommand \@sanitize [0]{\begingroup\catcode`\&12\catcode`\#12\relax}%
\@ifxundefined \pdfoutput {\@firstoftwo}{%
 \@ifnum{\z@=\pdfoutput}{\@firstoftwo}{\@secondoftwo}%
}{%
 \providecommand\@@startlink[1]{\leavevmode}%
 \providecommand\@@endlink[0]{}%
}{%
 \providecommand\@@startlink[1]{%
  \leavevmode
  \pdfstartlink
   attr{/Border[0 0 1 ]/H/I/C[0 1 1]}%
   user{/Subtype/Link/A<</Type/Action/S/URI/URI(#1)>>}%
  \relax
 }%
 \providecommand\@@endlink[0]{\pdfendlink}%
}%
\providecommand \url  [0]{\begingroup\@sanitize \@url }%
\providecommand \@url [1]{\endgroup\@href {#1}{\urlprefix}}%
\providecommand \urlprefix [0]{URL }%
\providecommand \Eprint[0]{\href }%
\@ifxundefined \urlstyle {%
  \providecommand \doi [1]{doi:\discretionary{}{}{}#1}%
}{%
  \providecommand \doi [0]{doi:\discretionary{}{}{}\begingroup
  \urlstyle{rm}\Url }%
}%
\providecommand \doibase [0]{http://dx.doi.org/}%
\providecommand \Doi[1]{\href{\doibase#1}}%
\providecommand \bibAnnote [3]{%
  \BibitemShut{#1}%
  \begin{quotation}\noindent
    \textsc{Key:}\ #2\\\textsc{Annotation:}\ #3%
  \end{quotation}%
}%
\providecommand \bibAnnoteFile [2]{%
  \IfFileExists{#2}{\bibAnnote {#1} {#2} {\input{#2}}}{}%
}%
\providecommand \typeout [0]{\immediate \write \m@ne }%
\providecommand \selectlanguage [0]{\@gobble}%
\providecommand \bibinfo [0]{\@secondoftwo}%
\providecommand \bibfield [0]{\@secondoftwo}%
\providecommand \translation [1]{[#1]}%
\providecommand \BibitemOpen[0]{}%
\providecommand \bibitemStop [0]{}%
\providecommand \bibitemNoStop [0]{.\EOS\space}%
\providecommand \EOS [0]{\spacefactor3000\relax}%
\providecommand \BibitemShut [1]{\csname bibitem#1\endcsname}%
\bibitem{chern2007prl}%
  \BibitemOpen
  \bibfield{author}{%
  \bibinfo {author} {\bibfnamefont{C.-H.}\ \bibnamefont{Chern}}\ and\ \bibinfo
  {author} {\bibfnamefont{D.-H.}\ \bibnamefont{Lee}},\ }%
  \bibfield{journal}{%
  \bibinfo {journal} {Phys. Rev. Lett.}\ }%
  \textbf{\bibinfo {volume} {98}},\ \bibinfo {pages} {066804} (\bibinfo {year}
  {2007})%
  \bibAnnoteFile{NoStop}{chern2007prl}%
\bibitem{trugman1985PRB}%
  \BibitemOpen
  \bibfield{author}{%
  \bibinfo {author} {\bibfnamefont{S.~A.}\ \bibnamefont{Trugman}}\ and\
  \bibinfo {author} {\bibfnamefont{S.}~\bibnamefont{Kivelson}},\ }%
  \bibfield{journal}{%
  \bibinfo {journal} {Phys. Rev. B}\ }%
  \textbf{\bibinfo {volume} {31}},\ \bibinfo {pages} {5280} (\bibinfo {year}
  {1985})%
  \bibAnnoteFile{NoStop}{trugman1985PRB}%
\bibitem{chern2010prb}%
  \BibitemOpen
  \bibfield{author}{%
  \bibinfo {author} {\bibfnamefont{C.-H.}\ \bibnamefont{Chern}},\ }%
  \bibfield{journal}{%
  \bibinfo {journal} {Phys. Rev. B}\ }%
  \textbf{\bibinfo {volume} {81}},\ \bibinfo {pages} {115123} (\bibinfo {year}
  {2010})%
  \bibAnnoteFile{NoStop}{chern2010prb}%
\bibitem{bernevig2002ANN}%
  \BibitemOpen
  \bibfield{author}{%
  \bibinfo {author} {\bibfnamefont{B.~A.}\ \bibnamefont{Bernevig}}, \bibinfo
  {author} {\bibfnamefont{C.-H.}\ \bibnamefont{Chern}}, \bibinfo {author}
  {\bibfnamefont{J.-P.}\ \bibnamefont{Hu}}, \bibinfo {author}
  {\bibfnamefont{N.}~\bibnamefont{Toumbas}},\ and\ \bibinfo {author}
  {\bibfnamefont{S.-C.}\ \bibnamefont{Zhang}},\ }%
  \bibfield{journal}{%
  \bibinfo {journal} {Annals of Physics}\ }%
  \textbf{\bibinfo {volume} {300}},\ \bibinfo {pages} {185} (\bibinfo {year}
  {2002})%
  \bibAnnoteFile{NoStop}{bernevig2002ANN}%
\bibitem{Lee2004PRL}%
  \BibitemOpen
  \bibfield{author}{%
  \bibinfo {author} {\bibfnamefont{D.-H.}\ \bibnamefont{Lee}}\ and\ \bibinfo
  {author} {\bibfnamefont{J.}~\bibnamefont{Leinaas}},\ }%
  \bibfield{journal}{%
  \bibinfo {journal} {Phys. Rev. Lett.}\ }%
  \textbf{\bibinfo {volume} {92}},\ \bibinfo {pages} {096401} (\bibinfo {year}
  {2004})%
  \bibAnnoteFile{NoStop}{Lee2004PRL}%
\bibitem{Seidel2005}%
  \BibitemOpen
  \bibfield{author}{%
  \bibinfo {author} {\bibfnamefont{A.}~\bibnamefont{Seidel}}, \bibinfo {author}
  {\bibfnamefont{H.}~\bibnamefont{Fu}}, \bibinfo {author}
  {\bibfnamefont{D.-H.}\ \bibnamefont{Lee}}, \bibinfo {author}
  {\bibfnamefont{J.~M.}\ \bibnamefont{Leinaas}},\ and\ \bibinfo {author}
  {\bibfnamefont{J.}~\bibnamefont{Moore}},\ }%
  \bibfield{journal}{%
  \bibinfo {journal} {cond-mat/0509071}}%
   (\bibinfo {year} {2005})%
  \bibAnnoteFile{NoStop}{Seidel2005}%
\bibitem{oshikawa2000PRL}%
  \BibitemOpen
  \bibfield{author}{%
  \bibinfo {author} {\bibfnamefont{M.}~\bibnamefont{Oshikawa}},\ }%
  \bibfield{journal}{%
  \bibinfo {journal} {Phys. Rev. Lett.}\ }%
  \textbf{\bibinfo {volume} {84}},\ \bibinfo {pages} {1535} (\bibinfo {year}
  {2000})%
  \bibAnnoteFile{NoStop}{oshikawa2000PRL}%
\end{thebibliography}
%

\end{document}